\documentclass{iopjournal}

\usepackage[separate-uncertainty=true, per-mode=repeated-symbol]{siunitx}

\usepackage[natbib,
    doi=true,
    sorting=none,
    isbn=false,
    url=false,
    eprint=false,
    maxnames=20,
    minnames=1,
    maxcitenames=1,
    mincitenames=1,
    giveninits=false,
    terseinits=true
]{biblatex}

\usepackage[version=4]{mhchem}
\usepackage{algorithm}
\usepackage{algpseudocode}
\usepackage{wrapfig} 
\usepackage{ragged2e}
\usepackage{here}
\AtBeginDocument{\justifying}

\usepackage{calc}

\begin{document}

\title{Too Big, Too Small, Too \ce{O2}: The Pandoro Effect from Oxygen Gradients in Tomographic Volumetric Additive Manufacturing}

\author{\makebox[0pt][l]{Felix Wechsler}\makebox[4pt][l]{\hspace{-0.1cm}Qianyi Zhang}, \hspace{-0.45cm}Riccardo Rizzo, Riccardo Rizzo$^{1, \dag, *}$\orcid{0000-0001-8297-6776}, Felix Wechsler$^{1, \dag, *}$\orcid{0000-0001-8763-2948}, Qianyi Zhang$^{1,\dag, *}$\orcid{0000-0001-8210-6893} and\\Christophe Moser$^{1,*}$\orcid{ 0000-0002-2078-0273}}

\fancyhead[R]{R. Rizzo, F. Wechsler, Q. Zhang, C. Moser}
\fancyhead[L]{The Pandoro effect in TVAM}

\affil{$^1$Laboratory of Applied Photonics Devices, School of Engineering, Ecole Polytechnique Fédérale de Lausanne, Switzerland}

\affil{$^\dag$These authors contributed equally and were sorted alphabetically.}

\affil{$^*$Author to whom any correspondence should be addressed.}

\email{riccardo.rizzo@epfl.ch, pandoro@felixwechsler.science, qianyi.zhang@epfl.ch, christophe.moser@epfl.ch}


\begin{abstract}
\justifying
Tomographic Volumetric Additive Manufacturing (TVAM) enables rapid, layerless biofabrication; however, its application to thermoreversible hydrogels is often compromised by complex chemical kinetics. 
In this study, we identify and characterize a recurrent printing artifact - termed the \textit{Pandoro effect} - manifesting as a truncated-cone distortion caused by premature polymerization at the vial bottom and inhibition at the top. 
We demonstrate that this phenomenon originates from a vertical oxygen gradient driven by the thermal hysteresis of resin preparation: heating depletes dissolved oxygen, while subsequent cooling induces diffusion-limited re-oxygenation from the air–resin interface.

To mitigate this, we present a multi-tiered strategy. 
First, we introduce a coupled ray-optical and photochemical optimization model that rigorously accounts for spatially heterogeneous inhibitor concentrations. 
Unlike conventional threshold-based approaches, this differentiable framework explicitly simulates the spatiotemporal reaction-diffusion dynamics of oxygen depletion, allowing the inverse solver to predictively compensate for local inhibition gradients. 
Complementing this algorithmic correction, we validate two process-based interventions: the elimination of the air–resin interface and the control of headspace atmosphere. 
We demonstrate that these strategies effectively suppress the \textit{Pandoro effect}, and are compatible with cell-laden resins. 
This work establishes guidelines for reproducible volumetric bioprinting and expands our open-source Dr.TVAM platform with advanced polymerization modeling capabilities.
\end{abstract}

\section{Introduction}
    Light-based fabrication technologies are rapidly emerging as powerful and versatile tools in tissue engineering and 3D bioprinting \cite{Lee_2021, levato_light-based_2023}. Among them, tomographic volumetric additive manufacturing (TVAM) \cite{Kelly_Bhattacharya_Heidari_Shusteff_Spadaccini_Taylor_2019, Bernal_Delrot_Loterie_Li_Malda_Moser_Levato_2019, Loterie_Delrot_Moser_2020} has attracted increasing attention for its unique ability to generate centimeter-scale constructs with high three-dimensional complexity within seconds (\autoref{fig:introfig}A). This layerless approach enables rapid, volumetric photopolymerization, overcoming many of the time and resolution constraints associated with conventional layer-by-layer printing strategies.
    
\subsection{Materials}
    A variety of biocompatible resin formulations have been adapted for TVAM \cite{madrid-wolff_review_2023}, with gelatin-based systems—such as gelatin methacryloyl (GelMA) and thiol–norbornene-functionalized gelatin (Gel-SH / NB)\cite{Rizzo_2021}—being among the most widely used. Gelatin, a denatured derivative of collagen, is extensively used in biofabrication due to its biocompatibility, biodegradability, intrinsic bioactivity, and ease of functionalization. Importantly, its reversible thermal gelation provides a distinct advantage in TVAM. When processed in the physically crosslinked (cooled) state, gelatin-based resins suppress object sedimentation during printing and enhance structural fidelity, thereby improving geometric accuracy.

  \subsection{Pattern optimization for TVAM}
        A key challenge in TVAM is determining which projection patterns are required to achieve successful printing under varying optical and chemical conditions.
        The evolution of pattern optimization in TVAM has transitioned from simple geometric modeling (e.g., Radon transform) \cite{Kelly_Bhattacharya_Heidari_Shusteff_Spadaccini_Taylor_2019, Loterie_Delrot_Moser_2020} to complex, physically accurate inverse problems. Early strategies relied on thresholding models, where the optimization objective was to ensure that void voxels remained below a polymerization limit while object voxels exceeded a curing threshold \cite{Kelly_Bhattacharya_Heidari_Shusteff_Spadaccini_Taylor_2019,Loterie_Delrot_Moser_2020, Rackson_Champley_Toombs_Fong_Bansal_Taylor_Shusteff_McLeod_2021}. To improve print fidelity, these heuristics were generalized into inverse modeling frameworks employing flexible loss functions \cite{BHATTACHARYA2021102299, Li_Toombs_Taylor_Wallin_2024, Nicolet_Wechsler_Madrid-Wolff_Moser_Jakob_2024}.
        
        As the demand for optical precision increased, forward models incorporated ray tracing to account for refraction and non-collimated illumination sources \cite{Webber_Zhang_Picard_Boisvert_Paquet_Orth_2023}. 
        While wave optics formulations for TVAM have been theoretically proposed to enhance resolution limits \cite{Wechsler_Gigli_Madrid-Wolff_Moser_2024, Li:24}, a wave-optics approach has been experimentally demonstrated only in a TVAM-similar printer \cite{Wang_Ma_Niu_Xiong_Zhang_Zhang_Chen_Wei_Fang_Wu_et}. 
        These ray-optical considerations have been unified by our group under inverse rendering techniques \cite{Nicolet_Wechsler_Madrid-Wolff_Moser_Jakob_2024} capable of modeling complex light-matter interactions - including attenuation, refraction, scattering, and reflection - across various scenarios, such as multi-material overprinting \cite{Wechsler_Sgarminato_Rizzo_Nicolet_Jakob_Moser_2025}.
        
        Despite these optical advancements, accounting for chemical kinetics remains a critical challenge. Variations in oxygen concentration during resin preparation are observed \cite{Kruse_Pellizzon_Salajeghe_Spangenberg_Lucklum_Islam_2026} and significantly influence polymerization dynamics \cite{Zhang_deHaan_Houlahan_Sampson_Webber_Orth_Lacelle_Gaburici_Lam_Deore_et, zhanginhibitor, THIJSSEN2024106096}. While dynamic corrections for oxygen diffusion during printing have been implemented \cite{Orth_Webber_Zhang_Sampson}, they assume a spatially homogeneous initial oxygen concentration. Recently, explicit inhibitor (namely oxygen and TEMPO) modeling was introduced in single-view holographic volumetric printing (SHVAM) \cite{Wechsler_Rizzo_Moser_2026}. This model enables the integration of spatially varying initial oxygen concentrations into the optimization process, a capability that is exploited here to address local inhibition concentrations.

    \subsection{The problem background}
\begin{figure}[h]
        \centering
        \includegraphics[width=1\linewidth]{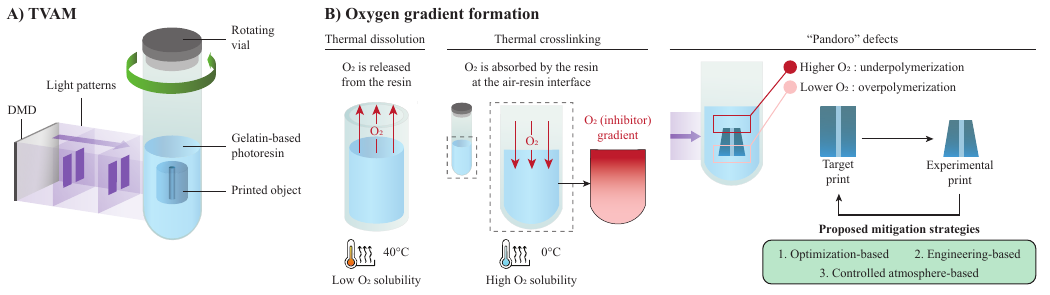}
        \caption{A) Schematic of tomographic volumetric additive manufacturing (TVAM) process. B) Schematic showing the formation of oxygen gradient following vial preparation with thermoreversible gelatin-based resin, and appearance of the Pandoro-like defects upon printing.}
        \label{fig:introfig}
    \end{figure}

    Sporadic but recurrent printing artifacts have been observed by the community, particularly in thermoreversible gelatin-based resins used in TVAM; however, these artifacts have not yet been systematically documented in the literature. They appear to result from non-uniform polymerization along the vertical axis of the resin volume. When present, the phenomenon consistently manifests itself as premature polymerization onset at the bottom of the vial and delayed polymerization at the top. Such behavior suggests the presence of a vertical gradient influencing polymerization kinetics, potentially arising from either non-uniform light distribution or spatial variations in resin composition. After excluding light inhomogeneity as the primary cause through careful optical alignment, we investigated the chemical origin of this effect.
    Oxygen is a well-established inhibitor of free-radical polymerization, as it readily reacts with initiating and propagating radicals generated upon light exposure. 
    In standard photoresins that do not undergo thermal cycling, dissolved oxygen typically remains at equilibrium with the surrounding atmosphere, resulting in a relatively uniform inhibitor concentration. 
    However, gelatin-based resins require a heating step to ensure complete dissolution, which reduces dissolved oxygen levels according to Henry’s law constants.
    Upon subsequent cooling in the printing vial to induce thermal gelation, the air–resin equilibrium is perturbed. During this stage, oxygen diffuses from the headspace into the resin, establishing a time-dependent vertical gradient (\autoref{fig:introfig}B). 
    The resulting higher oxygen concentration near the air–resin interface leads to spatially heterogeneous inhibition of radical polymerization.

    \subsection{Our contribution}
    Here, we systematically investigate and rationalize this phenomenon - termed the \textit{Pandoro effect} due to the characteristic truncated-cone geometry of failed prints. Through a combination of experimental measurements and simulations, we validate the oxygen-gradient hypothesis as the underlying mechanism. Building on this mechanistic understanding, we propose and demonstrate three separate mitigation strategies: 
    \begin{itemize}
        \item[i)] software-based correction via oxygen-gradient- and diffusion-informed pattern optimization,
        \item[ii)] engineering-based elimination of the air–resin interface,
        \item[iii)] atmosphere-controlled printing.
    \end{itemize}
    These approaches are validated in both cell-free and cell-laden systems.
    Collectively, this study elucidates and resolves a recurrent limitation in gelatin-based TVAM, establishing robust guidelines for reliable volumetric bioprinting with thermoreversible bioresins. Importantly, it also expands the photochemical framework of the open-source optical platform Dr.TVAM, enabling broader and more reproducible implementation of this versatile fabrication technology.

\section{Results and discussion}
\subsection{The Pandoro effect: experimental validation}
    The \textit{Pandoro effect} originates from the temperature-dependent solubility and diffusion of oxygen in thermoreversible resins. Using Gel-MA as a representative system, the resin is first heated to approximately \SI{40}{\celsius} for about \SI{1}{\hour} to ensure complete dissolution and homogeneous mixing. 
    At a given temperature, Henry’s law relates dissolved $\ce{O2}$ to its partial pressure; as temperature rises, the decreasing Henry’s law constant reduces $\ce{O2}$ solubility in water.
    Taking oxygen solubility in water as a reference, the equilibrium concentration decreases from $\approx \SI{8.7}{\milli\gram\per\liter}$ at \SI{22}{\celsius} to $\approx \SI{6.4}{\milli\gram\per\liter}$ at \SI{40}{\celsius} \cite{oxygenconc}. Consequently, during heating and mixing, a substantial fraction of dissolved oxygen is expected to outgas from the resin, leaving it in an oxygen-depleted state relative to ambient equilibrium conditions.
    After preparation, the warm resin is poured into a glass vial and rapidly cooled on ice to induce thermal gelation. The temperature quickly drops to approximately \SI{0}{\celsius}, and the resin is then stored for a certain varying time prior to printing, ranging from minutes to several hours. Immediately after cooling, the dissolved oxygen concentration within the bulk is spatially uniform but undersaturated with respect to its new equilibrium value at \SI{0}{\celsius} ($\approx \SI{14.6}{\milli\gram\per\liter}$) \cite{oxygenconc}.
    As a result, oxygen diffuses from the air–resin interface into the bulk hydrogel in order to re-establish equilibrium. Because the resin is no longer mixed and is physically gelled, oxygen transport occurs exclusively by diffusion. This diffusion-limited process generates a vertical oxygen concentration gradient, characterized by higher oxygen levels near the surface and lower concentrations toward the bottom of the vial.
    
\begin{figure}[h]
        \centering
        \includegraphics[width=1\linewidth]{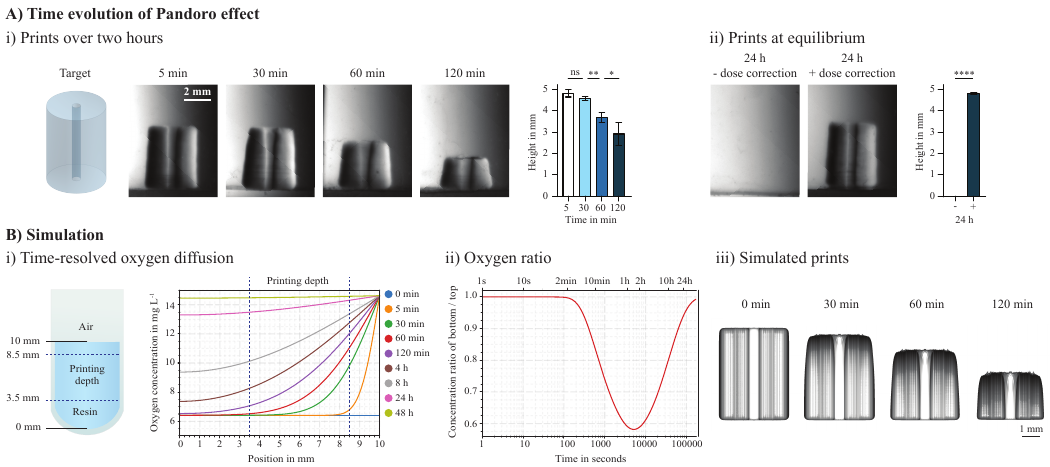}
        \caption{A-i) Representative prints and corresponding measured heights (n = 3) of the target model obtained 5, 30, 60, and \SI{120}{\minute} after vial preparation, illustrating the progressive development of the \textit{Pandoro effect} and ii) prints and measured heights at equilibrium (\SI{24}{\hour}) are shown without dose correction and after light-dose adjustment. B) Numerical simulation of the time-dependent oxygen gradient i) reveals different concentrations within the vial and ii) varying concentration differences between the top and the bottom of the structure, iii) simulated prints match the experimental trends.}
        \label{fig:pandorotime}
    \end{figure}

    Given the well-established inhibitory role of oxygen in free-radical photopolymerization - through scavenging of initiating and propagating radical species - this gradient produces spatially heterogeneous polymerization kinetics during printing. Regions near the air interface experience stronger inhibition and delayed polymerization onset, whereas deeper regions polymerize earlier due to their lower oxygen content. As shown in \autoref{fig:pandorotime}A-i, TVAM printing of a cylindrical model with a central hollow channel reveals the characteristic, time-dependent geometric distortion termed the \textit{Pandoro effect}. When printing is performed \SI{5}{\minute} after cooling, the resulting structures reach the intended height ($\approx \SI{5}{\mm}$), indicating minimal gradient formation. In contrast, printing after \SI{30}{\minute}, \SI{60}{\minute} or \SI{2}{\hour} yields shorter and visibly deformed constructs, consistent with elevated oxygen concentration in the upper region of the resin volume.
    After sufficiently long storage times (on the order of \SI{24}{\hour}), oxygen diffusion leads to an almost uniform concentration throughout the resin. At this stage, the overall inhibitor concentration has increased from $\approx \SI{6.4}{\milli\gram\per\liter}$ (saturation at \SI{40}{\celsius}) to $\approx \SI{14.6}{\milli\gram\per\liter}$ (saturation at \SI{0}{\celsius}). Under these conditions, the light dose previously sufficient during the first 2 hours is no longer adequate to initiate polymerization (\autoref{fig:pandorotime}A-ii). However, increasing the delivered light dose to overcome the higher oxygen concentration enables defect-free prints. Once equilibrium is reached and the vertical oxygen gradient disappears, the \textit{Pandoro effect} is no longer observed.
    
    To determine whether the \textit{Pandoro effect} could be eliminated by avoiding oxygen inhibition altogether, we also evaluated a thiol–norbornene gelatin system (Gel-SH/NB). Thiol–ene photochemistry is commonly described as oxygen-insensitive. Nevertheless, a pronounced \textit{Pandoro effect} was still observed with this resin (see \textit{Supplementary Information} - \autoref{fig:oxygen_insensitive}).
    The term oxygen-insensitive can be misleading. Compared to chain-growth methacryloyl polymerization (e.g., Gel-MA), thiol–ene reactions are indeed more tolerant to oxygen. In conventional (meth)acryloyl systems, oxygen rapidly reacts with carbon-centered radicals to form relatively unreactive peroxy radicals, effectively depleting the initiating species and leading to significant inhibition. In thiol–ene systems, however, these peroxy radicals can (at a slower rate) abstract a hydrogen atom from thiol groups, regenerating reactive thiyl radicals that continue the step-growth propagation (see \textit{Supplementary Information} - \autoref{fig:oxygen_insensitive}). As a result, oxygen does not fully suppress polymerization but rather delays it.
    Importantly, oxygen still competes for primary photoinitiator-derived radicals and transiently reduces the effective radical concentration. Therefore, spatial differences in dissolved oxygen concentration will still translate into spatial variations in polymerization kinetics. Even in thiol–ene systems, an oxygen gradient can thus produce heterogeneous crosslinking rates and ultimately Pandoro-like geometric distortions. From a theoretical standpoint, truly oxygen insensitive resins would be photoinitiator and radical-free systems, such as those based on photouncaging \cite{Rizzo_FreeRadical_2022, Kaneko_2026}, and photolysis\cite{Rizzo_oNBA_2025, guo_light-induced_2020}.  
    
    Concomitant with the appearance of the \textit{Pandoro effect}, we observed that the channel geometry is no longer straight but instead exhibits a funnel-like shape, with the diameter varying along its height (\autoref{fig:pandorotime}A-i). This distortion can be attributed to oxygen diffusion occurring during the printing process, as further discussed in the \textit{Supplementary Information} (\autoref{sec:tornado}).

\subsection{The Pandoro effect: numerical modeling of oxygen diffusion}
\label{sec:forwardmodel}
    
    At a concentration of \SI{7}{\percent}, Gel-MA forms a highly dilute hydrogel precursor; consequently, the transport properties of dissolved oxygen are predominantly governed by the aqueous solvent. We therefore approximate these properties using established literature values for water.
    
    During the initial heating phase to \SI{40}{\celsius}, mechanical mixing ensures the resin rapidly reaches the equilibrium oxygen concentration associated with this temperature, bypassing the slow timescale of pure diffusion. At \SI{0}{\celsius}, the equilibrium oxygen concentration in water is $c_{\SI{0}{\celsius}} = \SI{14.621}{\mg\per\L}$, whereas at \SI{40}{\celsius} it is significantly lower, $c_{\SI{40}{\celsius}} = \SI{6.421}{\mg\per\L}$ \cite{oxygenconc}.
    To estimate the oxygen diffusion coefficient in the Gel-MA resin at \SI{0}{\celsius}, we utilize the experimentally fitted equation reported by \citet{Han_Bartels_1996}, which yields $D \approx \SI{960}{\micro\meter\squared\per\s}$.
    For the geometric model, we assume a flat air-resin interface and a constant resin height of \SI{10}{\mm} within the vial.

    Upon cooling the vial to \SI{0}{\celsius}, the initial bulk oxygen concentration corresponds to $c_{\SI{40}{\celsius}}$. However, governed by Henry's law, the saturation concentration at the open surface immediately rises to $c_{\SI{0}{\celsius}}$.
    We assume the vial to be a finite cylinder filled with a certain resin level and hence we reduce the problem to 1D diffusion along the vertical axis. We solve Fick's second law along the vial height $h$:
    \begin{equation}
        \frac{\partial c}{\partial t} = D\frac{\partial^2 c}{\partial h^2},
    \end{equation}
    where $c(t, h)$ is the concentration, $D$ is the diffusion coefficient, and $t$ is time. We solve this equation using an explicit finite-difference scheme.
    While the analytical solution involving the complementary error function assumes a semi-infinite domain, it served as a validation metric for our numerical results, which account for the finite container height. The boundary condition at the surface is fixed at $c(t, h=\SI{10}{\mm}) = c_{\SI{0}{\celsius}}$ for all $t > 0$ and additionally the initial concentration is $c(0, h) = c_{\SI{40}{\celsius}}$.
    
    The resulting oxygen concentration profiles are illustrated in \autoref{fig:pandorotime}B-i. The TVAM patterns are defined within a printing range of $h=\SI{3.5}{\mm}$ to $h=\SI{8.5}{\mm}$ (\SI{1.5}{\milli\meter} from the air-resin interface). We observe that starting at \SI{30}{\min}, a significant vertical gradient develops. After \SI{24}{\hour} of diffusion, the system approaches equilibrium, with the concentration difference between the top and bottom of the print volume diminishing to less than \SI{10}{\percent}.
    
    To quantify the severity of the \textit{Pandoro effect}, we analyze the oxygen concentration ratio between the bottom and top of the printable area, $\frac{c(t, h=\SI{3.5}{\mm})}{c(t, h=\SI{8.5}{\mm})}$, as shown in \autoref{fig:pandorotime}B-ii. In the system considered here, the \textit{Pandoro effect} peaks at approximately \SI{90}{\minute}, where the oxygen concentration at the bottom is less than \SI{60}{\percent} of the concentration at the top. While TVAM is robust (determined by the volumetric energy dose separation between void and object regions) to minor deviations in inhibitor concentration, gradients of this magnitude can compromise print fidelity as demonstrated in \autoref{fig:pandorotime}A.

    \subsection{Optimization-based mitigation strategy}
        \label{sec:physicsforward}
        In principle, knowledge of spatially varying oxygen concentrations allows for the correction of TVAM patterns to compensate for local inhibition by delivering higher light doses to regions with elevated oxygen levels. However, conventional pattern optimization approaches typically neglect these chemical effects, relying instead on simplified polymerization threshold models that assume a homogeneous inhibitor concentration. 
        In this work, we introduce a coupled differentiable ray-optical and photochemical optimization framework for TVAM, wherein pattern optimization is directly coupled to the local inhibitor concentration. This approach builds upon the theoretical foundations established by \citet{Wechsler_Rizzo_Moser_2026}.

    \begin{wrapfigure}{r}{0.5\textwidth}
        \centering
        \includegraphics[width=\linewidth]{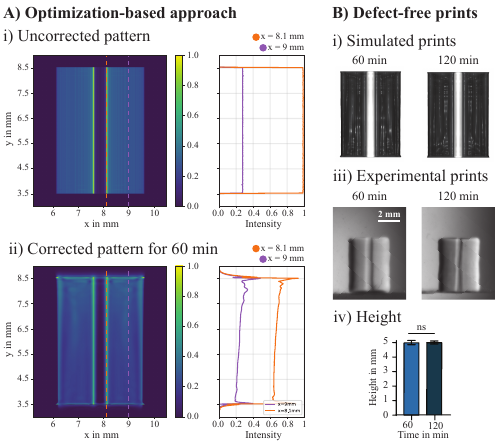}
        \caption{A) Comparison between patterns i) without and ii) with correction compensating for oxygen gradients and related diffusion. The close-up highlights higher intensities delivered to regions with higher oxygen concentrations. B) Defect-free i) simulated, and ii) experimental prints with iii) correspinding measured heights (n = 3) at 60 and \SI{120}{\minute} using gradient-aware patterns.}
        \label{fig:softwaremitigation}
    \end{wrapfigure}

        Unlike standard thresholding models, which optimize patterns merely to cross an intensity threshold for object voxels while remaining below it for void voxels \cite{Rackson_Champley_Toombs_Fong_Bansal_Taylor_Shusteff_McLeod_2021, BHATTACHARYA2021102299, Nicolet_Wechsler_Madrid-Wolff_Moser_Jakob_2024}, our framework explicitly models two state variables: the volumetric oxygen concentration and the polymerization dose. 
        Although the complete chemical kinetics of TVAM are complex \cite{weisgraber2023virtual}, we employ a tractable model based on reaction rates following \citet{Wechsler_Rizzo_Moser_2026}. Based on typical reaction-rate constants, this model assumes that: 
        i) absorbed photons generate radicals; 
        ii) these radicals are immediately quenched by the local inhibitor \cite{Zhang_deHaan_Houlahan_Sampson_Webber_Orth_Lacelle_Gaburici_Lam_Deore_et}; 
        iii) polymerization-initiating dose accumulates only after the local inhibitor has been fully depleted.
        
        Concurrent with the delivery of the spatially varying light dose and consequently local inhibitor depletion, inhibitor diffusion occurs within the resin volume, having a spatial influence on the order of a few hundred micrometers (calculated by the diffusion length $\sim 2 \sqrt{D \cdot t}$). 
        It is important to note that this inhibitor diffusion operates on significantly smaller scales than the \textit{Pandoro effect} discussed herein.
        Moreover, previous work \citet{Orth_Webber_Zhang_Sampson} addressed oxygen diffusion numerically during printing, their approach approximated inhibitor diffusion as light dose diffusion. This approximation is physically inexact and incompatible with the inhomogeneous oxygen concentrations present in our system. In contrast, we model inhibitor diffusion explicitly. Furthermore, 
        to optimize TVAM patterns under a spatially varying oxygen concentration, we need explicit tracking of inhibitor quantities as introduced in \autoref{sec:forwardmodel}.
        
        
        Qualitatively, the algorithm iterates as follows: The light dose for a single rotation is projected into the vial. Generated radicals are initially consumed by the local inhibitor; only upon local depletion of the inhibitor does the radical concentration contribute to the polymerization dose. Subsequently, a diffusion kernel is applied to the inhibitor concentration field via convolution. This process is repeated for the number of rotations specified by the operator.
        
        We implement this coupled ray-optical and photochemical optimization - which accounts for both initial oxygen heterogeneity and print-induced diffusion - within our open-source framework, Dr.TVAM \cite{Nicolet_Wechsler_Madrid-Wolff_Moser_Jakob_2024}. See \autoref{sec:algorithm}  and our open implementation for full algorithmic details.

    \subsubsection{Patterns without oxygen gradient correction}
        The prints shown in \autoref{fig:pandorotime} were generated using patterns optimized under the assumption of a homogeneous oxygen concentration ($c[\ce{O2}] = 1$). As illustrated in \autoref{fig:pandorotime}A-i, the experimentally measured height decreases as resin storage time increases. 
        Since oxygen diffuses to the top region of the print first, vertically homogeneous illumination is not enough to overcome inhibition and initiate polymerization.
        This trend is consistent with the simulated prints in \autoref{fig:pandorotime}B-iii, where different oxygen concentrations were initialized based on the physical simulation.
    
    \subsubsection{Patterns with oxygen gradient correction}
        By utilizing the simulated vertical oxygen concentration, we can initialize $c[\ce{O2}]$ to reflect the experimental conditions more accurately. The resulting pattern for the \SI{60}{\minute} storage time is displayed in \autoref{fig:softwaremitigation}A-ii. Notably, the optimization algorithm projects slightly higher intensities in the top region of the sample to overcome the elevated oxygen concentrations.
        Experimentally, projecting these corrected patterns into the resin yields consistent prints with no \textit{Pandoro effect}. Specifically, the object height is preserved across all trials, as confirmed by both the simulation \autoref{fig:softwaremitigation}B-i and \autoref{fig:softwaremitigation}B-ii-iii) experimental results.
        Additionally, modeling oxygen diffusion during the print time adds slightly more intensity on the edges of the object to counteract the edge blurring of the oxygen diffusion with the patterns (e.g. see $y=\SI{8.5}{\mm}$ in \textit{Supplementary Information} - \autoref{fig:optimization results}).

\subsection{Engineering-based mitigation strategy}
\begin{figure}[h]
        \centering
        \includegraphics[width=1\linewidth]{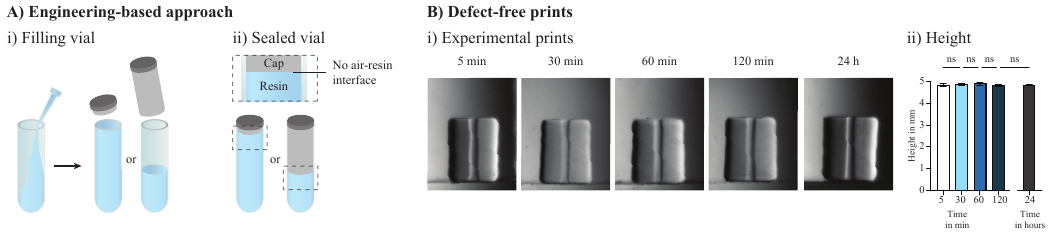}
        \caption {A) Schematic of vial preparation for air-interface removal showing use of standard caps (left) requiring high resin volume or custom caps (right) limiting resin consumption. B) Representative prints (i) and corresponding measured heights (n = 3) (ii) of the target model obtained at 5, 30, 60, 120 min and \SI{24}{\hour} after vial preparation, illustrating the absence of the \textit{Pandoro effect}.}
        \label{fig:fillingtotop}
    \end{figure}

    Because the oxygen gradient originates at the air–resin interface, a straightforward mitigation strategy is to eliminate this interface altogether. To test this hypothesis, we filled the vials completely and sealed them with a cap, ensuring the absence of air bubbles (air contains around 20 times more oxygen per volume than water so a single bubble can cause significantly spatially varying oxygen concentrations in the resin) and preventing any contact between the resin and the external atmosphere (\autoref{fig:fillingtotop}A). Under these conditions, defect-free prints were consistently obtained for up to \SI{24}{\hour} after vial preparation, indicating that no oxygen gradient developed over time (\autoref{fig:fillingtotop}B).
    
    This engineering-based solution provides a simple and effective way to suppress the \textit{Pandoro effect}. However, when standard laboratory vials are used, fully eliminating the headspace requires a substantially larger volume of resin than is strictly necessary for printing. In our setup, this approach demanded more than three times the resin volume otherwise required, leading to significant material waste. For applications involving valuable or cell-laden formulations, the use of custom-designed vials or dedicated caps (\autoref{fig:fillingtotop}A, \textit{Supplementary Information} - \autoref{fig:customcap}) is therefore recommended to minimize resin consumption while maintaining elimination of the air–resin interface.

\subsection{Controlled atmosphere-based mitigation strategy}
\begin{wrapfigure}{r}{0.5\textwidth}
        \centering
        \includegraphics[width=1\linewidth]{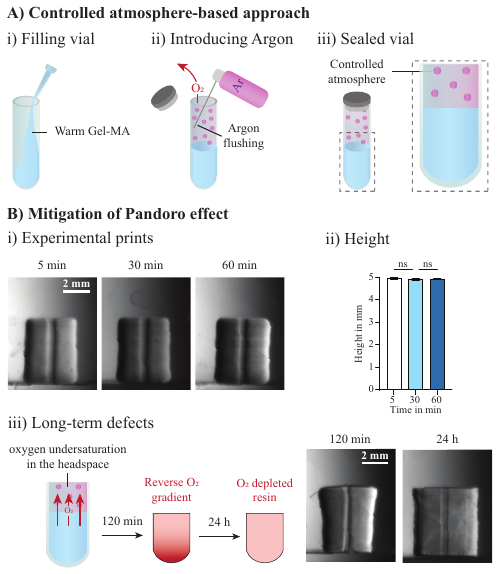}
        \caption{A) Schematic of vial preparation for controlled atmosphere-based approach. B) Representative defect-free prints (i) and corresponding measured heights (n = 3) (ii) of the target model obtained at 5, 30, and \SI{60}{\minute} afer vial preparation. The prints illustrate the absence of the \textit{Pandoro effect} within the first \SI{60}{\minute}, while a reverse \textit{Pandoro effect} (iii) and overpolymerization issues arise after \SI{120}{\minute} and \SI{24}{\hour}, respectively.}
        \label{fig:inertatmosphere}
    \end{wrapfigure}

    As discussed above, the \textit{Pandoro effect} stems from an imbalance in oxygen partial pressure between the air-filled headspace and the cooled resin. A direct mitigation strategy is therefore to control the oxygen concentration in the headspace to limit oxygen diffusion into the resin. When the resin is heated to  \SI{40}{\celsius}, the equilibrium oxygen concentration decreases to $\approx \SI{6.4}{\milli\gram\per\liter}$; upon cooling to \SI{0}{\celsius}, it increases to $\approx \SI{14.6}{\milli\gram\per\liter}$ (a $2.3$-fold increase). In principle, maintaining equilibrium and preventing gradient formation would require reducing the oxygen concentration in the headspace by the same factor. However, precise control of oxygen partial pressure is technically demanding and impractical for a rapid TVAM workflow.
    A more feasible approach is to partially reduce the headspace oxygen concentration to slow gradient formation and extend the defect-free printing window. To test this hypothesis, we flushed the vial headspace with \SI{99}{\percent} argon (commonly used for wine preservation) prior to sealing, thereby lowering the overall oxygen content (\autoref{fig:inertatmosphere}A). Under these conditions, no significant defects were observed within the first \SI{60}{\minute} (\autoref{fig:inertatmosphere}B), confirming delayed gradient formation.
                    
    Interestingly, after \SI{120}{\minute}, a \textit{reverse Pandoro effect} geometry emerged, characterized by overpolymerization at the top of the construct. This behavior likely results from undersaturation of oxygen molecules in the headspace during the air–argon mixing process, which drives oxygen out of the resin and establishes an inverted gradient with reduced inhibition near the surface. This interpretation is further supported by results at \SI{24}{\hour}: using the same light dose, constructs become significantly overpolymerized, with lateral dimensions limited only by the vial diameter. Because the headspace contains roughly two orders of magnitude more oxygen than the resin (the air volume inside the vial is larger than the resin volume and air contains more oxygen molecules per volume than liquids), small imbalances can deplete most of the dissolved oxygen, effectively eliminating inhibition.
    Although not a perfectly controlled solution, headspace flushing with inert, or virtually any non radical-inhibiting gas provides a simple and practical method to suppress the \textit{Pandoro effect} for up to \SI{60}{\minute}. This time window is well aligned with typical TVAM bioprinting workflows, where cell-laden resins are generally used within \SI{30}{\minute} of preparation (see \autoref{sec:bioprinting}).

\subsection{Bioprinting}
    \label{sec:bioprinting}
    Finally, we evaluated the proposed mitigation strategies under bioprinting conditions. In line with standard practice, and to avoid prolonged storage of cell-laden formulations ($\SI{2e6}{cells\per\milli\liter}$), printing was performed within \SI{30}{\minute} of vial preparation.

\begin{figure}[h]
        \centering
        \includegraphics[width=1\linewidth]{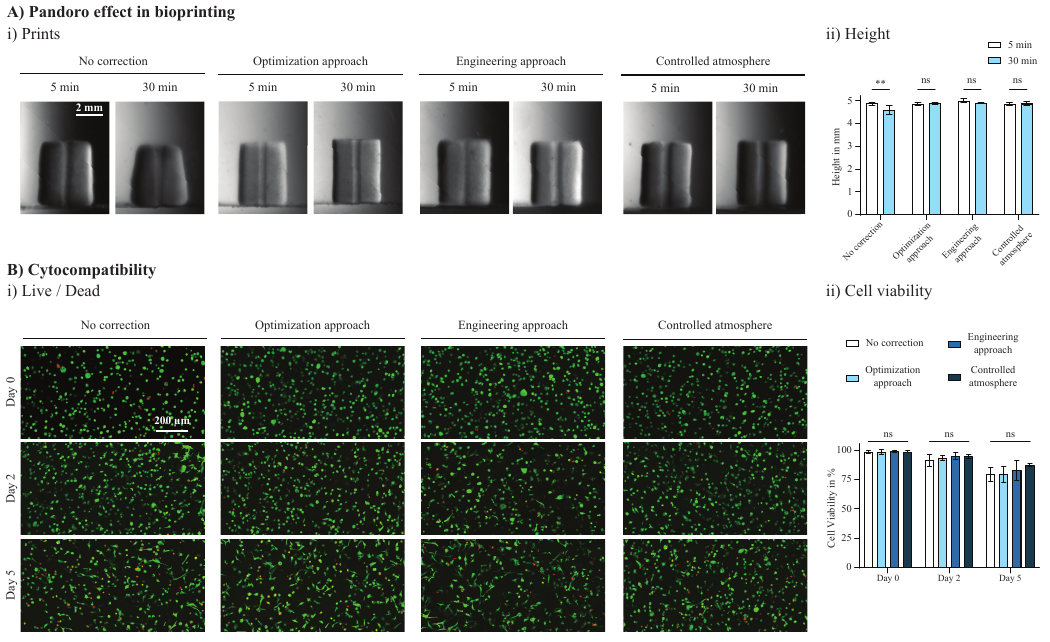}
        \caption{A) Experimental prints (i) and measured heights (n = 3) (ii) of the target model obtained with cell-laden formulation ($\SI{2e6}{cells\per\milli\liter}$) at 5 and \SI{30}{\minute} afer vial preparation. The prints show no evidence of the \textit{Pandoro effect} under any condition at \SI{5}{\minute}. In contrast, the non-mitigated prints fail after \SI{30}{\minute}, whereas all three proposed mitigation strategies result in defect-free prints. B) Representative Live/Dead confocal images after 0, 2 and 5 days of culture for all printing conditions (i) and viability results (ii) confirms no significant differences and overall high cytocompatibility of the proposed mitigation strategies.}
        \label{fig:bioresults}
    \end{figure}

    When printing was carried out after only \SI{5}{\minute} of cooling, all constructs reached the intended height and exhibited no visible \textit{Pandoro effect}-related distortions, even in the absence of mitigation (\autoref{fig:bioresults}A). This observation is consistent with the limited development of the oxygen gradient at early time points.
    
    In contrast, when printing was performed after \SI{30}{\minute}, a pronounced \textit{Pandoro effect} was observed in the non-mitigated condition, resulting in clear geometric deformation and reduced construct height. Notably, all three proposed mitigation strategies effectively suppressed this distortion, yielding structures that closely matched the target geometry (\autoref{fig:bioresults}A). These results confirm the robustness and practical relevance of the proposed solutions for cell-laden TVAM bioprinting workflows. Notably, for the engineering-based mitigation strategy, a custom cap was developed (see \textit{Supplementary Information} \autoref{fig:customcap}) to minimize the loss of valuable cell-laden resin.
    
     Importantly, none of the proposed mitigation strategies interfered with cell viability (\autoref{fig:bioresults}B). High viability ($> \SI{97}{\percent}$) is in fact reported upon printing (day 0), and after 2 days ($> \SI{90}{\percent}$ and 5 days $> \SI{80}{\percent}$ of culture, with no significant difference compared to the control, non-mitigated, condition.

\section{Conclusion and outlook}
    In this work, we have systematically characterized and mitigated the \textit{Pandoro effect}, a truncated-cone-shaped artifact in volumetric bioprinting arising from thermal history-driven oxygen gradients.
    By integrating physical experiments with advanced computational modeling, we established a comprehensive framework for reliable printing in thermoreversible hydrogels.
    
    Besides the systematic study of the \textit{Pandoro effect}, one primary contribution is the introduction of a coupled ray-optical and photochemical optimization model. 
    This framework goes beyond static thresholding to explicitly simulate both spatially varying inhibitor concentrations and dynamic diffusion processes during printing. 
    A key advantage of this approach is its robustness; we demonstrated that high-fidelity corrections can be achieved using standard literature values for aqueous diffusion, effectively eliminating the need for laborious, material-specific parameter calibration. 
    This renders our toolbox potentially applicable to a wide range of water-based bioresins currently used in the community. By incorporating physical effects into the open-source Dr.TVAM platform, this work significantly narrows the gap between computational prediction and experimental reality, advancing the fidelity and reproducibility of volumetric bioprinting.
    
    Complementing this algorithmic solution, we validated practical process interventions; specifically, the elimination of the air-resin interface and controlled headspace atmosphere.
    These strategies provide immediate, accessible solutions for standard workflows where computational optimization may not be required.

    Looking forward, the explicit modeling of reaction-diffusion kinetics opens new avenues for process refinement. 
    While this study focused on macroscopic uniformity, dynamic diffusion control holds the potential to enhance spatial resolution, as suggested by \citet{Orth_Webber_Zhang_Sampson}; we leave the systematic exploration of resolution limits with our coupled solver for future work. 
    Furthermore, in the context of bioprinting, the model could be expanded to account for the metabolic activity of encapsulated cells, which respire oxygen and dynamically alter the inhibitor landscape during storage and printing. 

Ultimately, although it was studied here in the context of gelatin-based resins and biofabrication applications, the \textit{Pandoro effect} is intrinsically linked to any photoresin that undergoes thermal cycles between preparation and printing. The mitigation strategies proposed in this work are particularly relevant for bioresins, which are ideally used shortly upon preparation, but they may also be applicable to other photoresins \cite{acrylate_temp} subjected to thermal cycling conditions.

\section{Materials and methods}

\subsection{Synthesis of gelatin-methacryloyl (Gel-MA)}
Gel-MA was synthesized following a modified version of a previously reported protocol by \citet{rizzo_tomographic_2026}. In brief, \SI{25}{\g} of type A gelatin (300 bloom, porcine skin) was dissolved in PBS at \SI{10}{\percent} (w/v) at \SI{45}{\celsius}. 
Once fully solubilized, \SI{15}{\mL} of methacrylic anhydride was added dropwise, and the reaction was allowed to proceed for \SI{1.5}{\hour}. The reaction mixture was then diluted twofold with prewarmed Milli-Q (mQ) water and centrifuged at \SI{4000}{rpm} for \SI{5}{\minute} to remove unreacted anhydride and acrylic acid. 
Subsequently, \SI{0.5}{\gram} of sodium chloride (\ce{NaCl}) was added to the supernatant, and the solution was dialyzed (MWCO 14 kDa, Roth AG) against mQ water for 4–5 days with frequent water replacement, followed by freeze-drying. 
The resulting lyophilized G-MA was stored at \SI{-20}{\celsius} until further use. 
The degree of substitution (DS) was determined by $^1$H-NMR in \ce{D2O}  and estimated at $\approx \SI{0.25}{mmol}$ of methacrylic groups per gram of gelatin, using 3-(trimethylsilyl)-1-propanesulfonic acid (DSS, 2H, \SI{0.65}{ppm}) as an internal standard and integrating the methacrylate vinyl protons (\SI{\sim 5.45}{ppm} and \SI{\sim 5.70}{ppm}).

\subsection{Synthesis of gelatin-thiol (Gel-SH)}
Gel-SH was prepared according to the method previously reported by \citet{rizzo_tomographic_2026}. Briefly, \SI{10}{\gram} of type A porcine gelatin (\SI{300}{bloom}) were dissolved in \SI{500}{\milli\liter} of \SI{150}{\milli M} MES buffer (pH 4.5). 
Then, \SI{0.48}{\gram} of 3,3-dithiobis(propionohydrazide) (DTPHY, Chemie Brunschwig AG) were added under stirring. After complete dissolution, \SI{1.5}{\gram} of 1-ethyl-3-(3-dimethylaminopropyl)carbodiimide (EDC, Roth AG) were introduced, and the reaction was allowed to proceed for \SI{24}{\hour}.
Subsequently, \SI{3.3}{\gram} of tris(2-carboxyethyl)phosphine (TCEP, Chemie Brunschwig AG) were added to reduce the disulfide bonds, and the reduction step was carried out for \SI{6}{\hour}. Afterward, \SI{1}{\gram} of sodium chloride (NaCl) was added, and the solution was dialyzed (MWCO \SI{14}{\kilo\dalton}, Roth AG) against acidified (pH 4) Milli-Q (mQ) water for 4–5 days with frequent water replacement. 
The resulting G-SH solution was sterile-filtered prior to freeze-drying. The lyophilized G-SH was stored at \SI{-20}{\celsius} until use. 
The degree of substitution (DS) was determined by $^1$H-NMR in \ce{D2O} and estimated at \SI{\sim 0.26}{\milli\mole} of thiol (SH) groups per gram of gelatin, using 3-(trimethylsilyl)-1-propanesulfonic acid (DSS, 2H, \SI{\sim 0.65}{ppm}) as an internal standard and integrating the hydrazide methylene proton signals (\SI{\sim 2.7}{ppm} and \SI{\sim 2.8}{ppm}).

\subsection{Synthesis of gelatin-norbornene (Gel-NB)}
G-NB was synthesized following the protocol previously reported by \citet{rizzo_tomographic_2026}. Briefly, \SI{20}{\gram} of type A porcine skin gelatin (\SI{300}{bloom}) were dissolved at \SI{10}{\percent} (w/v) in \SI{0.5}{M} carbonate–bicarbonate buffer (pH 9) at \SI{40}{\celsius}. After complete dissolution, \SI{0.4}{\gram} of cis-5-norbornene-endo-2,3-dicarboxylic anhydride (CA, Chemie Brunschwig AG) was added under vigorous stirring. Additional portions of \SI{0.4}{\gram} CA were introduced every \SI{10}{\minute}, reaching a total of \SI{2}{\gram}. \SI{20}{\minute} after the final addition, the reaction mixture was diluted twofold with prewarmed Milli-Q (mQ) water. Subsequently, \SI{2}{\gram} of sodium chloride (\ce{NaCl}) were added, and the solution was sterile-filtered (\SI{0.2}{\micro\meter}) and dialyzed (MWCO \SI{14}{\kilo\dalton}, Roth AG) against mQ water at \SI{30}{\celsius} for 4–5 days with frequent water changes. The product was then freeze-dried, and the resulting lyophilized G-NB was stored at \SI{-20}{\celsius} until use. The degree of substitution (DS) was quantified by $^1$H-NMR in \ce{D2O} and estimated at \SI{\sim 0.17}{\milli\mole} of norbornene (NB) groups per gram of gelatin, using 3-(trimethylsilyl)-1-propanesulfonic acid (DSS, 2H, \SI{\sim 0.65}{ppm}) as an internal standard and integrating the characteristic NB proton signals (\SIrange{\sim 6.4}{\sim 6.0}{ppm}).

\subsection{Cell culture}
Human foreskin fibroblasts (HFF-1 SCRC-1041, ATCC), were cultured in Dulbecco's Modified Eagle's Medium (DMEM) without phenol red, supplemented with antibiotic-antimycotic (Gibco), \SI{2}{\percent} L-glutamine (Gibco) and \SI{10}{\percent} fetal bovine serum (Gibco). Media was changed every other day, and cells were used at passage 8-14. For viability study, printed constructs were cultured under the same conditions in 48-well plates.

\subsection{Samples preparation}
Gel-MA and Gel-SH/NB were dissolved in PBS with \SI{0.05}{\percent} LAP at \SI{40}{\celsius}. For cell-free resins, Gel-MA and Gel-SH/NB were prepared at final concentration of \SI{7}{\percent} and \SI{5}{\percent}, respectively. When completely dissolved, the resin was filtered (\SI{0.44}{\micro\meter}) to eliminate scattering debris. The resin was then stirred at \SI{40}{\celsius} for \SI{1}{\hour} to reach oxygen equilibration. Then, \SI{200}{\micro\liter} were pipetted into glass vials (42775009 test tubes, Karl Hecht) and rapidly placed in ice for thermal gelation. For engineering-based solution, \SI{700}{\micro\liter} of resin were used to fill the test tube. For controlled atmosphere-based solution, argon (Preservintage, Pushback Ltd) was sprayed for \SI{2}{\second} in the filled vial prior to sealing with cap. For cell-laden resins, Gel-MA was dissolved at \SI{10}{\percent}, sterile filtered (\SI{0.2}{\micro\meter}) and diluted to \SI{7}{\percent} with a cell suspension in PBS under cell culture hood. The vials were prepared with  \SI{300}{\micro\liter} of cell-laden resin and custom caps (\autoref{fig:customcap}). For all conditions, after sealing the vials were readily placed in ice.

\subsection{TVAM setup}
TVAM was carried out using a custom-built setup, as previously reported. Briefly, a \SI{399}{\nano\meter}, \SI{8}{\watt} fiber-coupled high-power laser (CivilLaser, China) served as the light source. The laser beam was directed onto a high-speed digital micromirror device (DMD; VIS-7001, Vialux), which projected the patterned light onto the printing plane through a 4f optical system composed of \SI{150}{\milli\meter} and \SI{100}{\milli\meter} plano-convex lenses. A Fourier stop was incorporated to suppress higher diffraction orders. The resulting pixel size at the image plane was determined to be \SI{20.45}{\micro\meter}. Synchronization between the DMD and the high-precision rotation stage (X-RSW60C, Zaber) was achieved using an Arduino Nano Every.

\subsection{TVAM printing}
Pattern optimization was conducted with our in-house, open source Dr.TVAM optical framework as previously described on a NVIDIA L40S GPU or NVIDIA RTX 3060 \SI{12}{\giga\byte}. Absorption and refractive index of the resins were measured with a spectrophotometer (Cary 60, Agilent), and a digital refractometer (AR200, Reichert Inc.), respectively. Vials were taken from ice at different time points and secured onto the rotating stage. A camera system was employed to ensure precise vertical alignment of the vial, consistently positioning the projected pattern \SI{1.5}{\milli\meter} below the air–resin interface. An index matching water bath was adopted to limit artifacts from glass test tube imperfections. 
Upon printing, the vials were readily warmed up to \SI{40}{\celsius}, samples retrieved, and washed 2-3x with warm PBS. The printed constructs were transferred to a quartz cuvette (\SI{1}{\centi\meter} optical path length) and imaged using the system camera under oblique illumination for enhanced contrast. For cell-laden samples, the post-processing step was performed in a \SI{37}{\celsius} water bath and then under a sterile cell-culture hood.

\subsection{Confocal imaging}
Cells-embedded samples (n = 3 per time point) were taken on day 0, day 2, and day 5 and incubated in FluoroBrite DMEM supplemented with 1:2000 calcein-AM, 1:500 ethidium homodimer-1 and 1:1000 Hoechst 33342 for \SI{30}{\minute}. Imaging (\SI{200}{\micro\meter}, \SI{10}{\micro\meter} steps) was carried out using an inverted confocal microscope (Leica SP8) equipped with an HC PL Fluotar 10×/0.30 objective. Cell viability was quantified using ImageJ Analyze particle function.

\subsection{Statistical analysis}
Statistical analysis was performed using GraphPad Prism 10 and statistical
significance was determined using one-way Welch Anova with multiple comparisons, two-way Anova with multiple comparisons or unpaired Welch’s t-test.

\ack{We thank Viola Sgarminato for the initial branding of the effect under the name \textit{Panettone}. For correctness we adapted the name to \textit{Pandoro effect}. 
During the preparation of this work, the authors used generative AI tools to assist with improving source code and with editing the manuscript for grammar, spelling, and clarity (including this sentence).
We also value discussions with Eveline Mayner, Manuela Cedrún-Morales,  Deepika Sardana and Georges Wagnières on chemical mechanisms.}

\funding{This project has received funding from the Swiss National Science Foundation 2000-1-240074
under grant number 10007068 - \textit{Neural precision holographic volumetric additive
manufacturing} and Return CH Postdoc.Mobility
 P5R5$-$3\_235066 (R.R.).}

\roles{R.R., F.W., and Q.Z. contributed equally to this work and share first authorship.\\
R.R: Conceptualization, Data curation, Formal analysis, Funding acquisition, Investigation, Methodology, Resources, Validation, Visualization, Writing – original draft, Writing – review \& editing.\\
F.W.: Conceptualization, Data curation, Formal analysis, Investigation, Methodology, Resources, Software, Validation, Visualization, Writing – original draft,Writing – review \& editing.\\
Q.Z: Conceptualization, Data curation, Formal analysis,  Investigation, Methodology, Resources, Validation, Visualization, Writing – review \& editing.\\
C.M.: Funding acquisition, Project administration, Resources, Supervision, Writing – review \& editing
}

\suppdata{The relevant source code, mesh files and configuration files can be found here: \url{https://github.com/EPFL-LAPD/The-Pandoro-Effect-in-Tomographic-Volumetric-Additive-Manufacturing}}

\printbibliography

\clearpage
\section*{Supplementary Information}
\setcounter{figure}{0}
\renewcommand{\thefigure}{S\arabic{figure}}
\appendix

\section{Oxygen diffusion-related defects}
    \label{sec:tornado}
    Our uncorrected prints display channels with funnel-shaped distortion \autoref{fig:pandorotime}A-i (e.g., at \SI{60}{\minute}). This effect appears to be more pronounced at the top of the vial where oxygen concentration is higher. We hypothesize that this behavior arises because, in addition to the vertical oxygen gradient responsible for the \textit{Pandoro effect}, oxygen diffusion—and the associated inhibition during the printing process—was not accounted for. In regions where the oxygen concentration is higher (near the top of the vial), enhanced spatial diffusion following local depletion further delays polymerization.
    
\begin{figure}[H]
        \centering
        \includegraphics[width=0.5\linewidth]{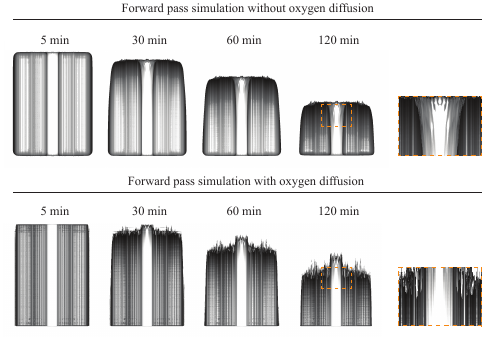}
        \caption{Forward pass simulations at \SI{5}{\minute}, \SI{30}{\minute}, \SI{60}{\minute}, and \SI{120}{\minute} with (top) and without (bottom) accounting for oxygen diffusion (and hence inhibition) during the print. Closeups show funnel-like defects in line with experimental observations (see \autoref{fig:pandorotime}A) present only if accounting for oxygen diffusion.}
        \label{fig:tornado}
    \end{figure}

    The underlying mechanism of this effect is also confirmed via simulation, clearly appearing when uncorrected patterns (not accounting for oxygen gradient nor diffusion) are used (\autoref{fig:pandorotime}B)iii. When we simulate the process with oxygen diffusion disabled during the print phase, the effect disappears, as shown in \autoref{fig:tornado}. This further proves that this effect is caused by oxygen diffusion during the exposure time.

    As discussed in detail in the manuscript and in \autoref{sec:algorithm} and \autoref{sec:opt_results}, in this work we implement corrections of TVAM patterns to compensate for vertical oxygen gradients responsible for the \textit{Pandoro effect}, as well as a photochemical optimization framework that accounts for local inhibition concentration and diffusion, thus solving also the funnel-like distortions.
    
\section{Coupled differentiable ray-optical and photochemical optimization}
\label{sec:algorithm}
    
    We employ a gradient-descent (L-BFGS optimizer \cite{Liu_Nocedal_1989}) optimization strategy underpinned by a physics-based forward model to generate high-fidelity projection patterns. The optical propagation is modeled using the Dr.TVAM framework \cite{Nicolet_Wechsler_Madrid-Wolff_Moser_Jakob_2024}, while the photochemical kinetics are modeled according to the principles introduced in \autoref{sec:forwardmodel} and \autoref{sec:physicsforward}.
    
    \subsection{Loss Function}
    The objective function is formulated to govern the interplay between polymer formation and oxygen inhibition. We define $\mathrm{ReLU}(x)=\max(0,x)$ and minimize the following loss:
    
    \begin{equation}
    \label{eq:loss}
    \begin{split}
    \mathcal{L} =
    &\;
    \phantom{+} w_\text{obj} \cdot \sum_{v \in \Omega_{\text{obj}}} 
    \underbrace{\left|\mathrm{ReLU}\!\left(T_P + c[\ce{O2}]_v - c[\ce{P}^*]_v \right)\right|^2}_{\substack{\text{Target conversion:} \\ \text{Dose must exceed threshold + deplete \ce{O2}}}}
    +
    \underbrace{\left|\mathrm{ReLU}\!\left(c[\ce{P}^*]_v - T_{\text{OP}}\right)\right|^2}_{\substack{\text{Over-curing penalty}}}
    \\
    &\; + 
    w_\text{void} \cdot  \sum_{v \in \Omega_{\text{void}}}
    \underbrace{\left|
    \mathrm{ReLU}\!\left(T_I + c[\ce{P}^*]_v - c[\ce{O2}]_v \right)
    \right|^2}_{\substack{\text{Void preservation:} \\ \ce{O2} \text{ must remain above safety margin}}} 
    \\
    &\; +
    w_\text{sparsity} \cdot \sum_{j, i} \underbrace{|P_{j,i}|^D}_{\text{Sparsity regularization}}
    .
    \end{split}
    \end{equation}
    
    Here, $c[\ce{O2}]_v$ and $c[\ce{P}^*]_v$ denote the local oxygen and polymer concentrations at voxel $v$, respectively, while $P_{j,i}$ represents the intensity modulation for projection angle $\alpha_j$ at detector pixel $i$.
    $\Omega_{\text{void}}$ and $\Omega_{\text{obj}}$ indicate the void and object voxels each.
    The exponent $D$ is an integer parameter chosen to enforce non sparse patterns (high energy efficiency). 
    The relative weights $w_\text{obj},  w_\text{void},  w_\text{sparsity}$ are chosen to be 1.
    The individual components of the loss function are defined as follows:
    
    \begin{itemize}
        \item \textbf{Target conversion:} This term ensures the effective polymerization dose $c[\ce{P}^*]$ surpasses the polymerization threshold $T_P$ (set to 0.4). The inclusion of $c[\ce{O2}]$ compels the optimizer to drive local oxygen concentrations toward zero, a prerequisite for radical propagation.
        \item \textbf{Over-curing penalty:} This constraint restricts the dose to an upper bound $T_{\text{OP}}$ (set to 0.62) to mitigate artifacts arising from excessive exposure.
        \item \textbf{Void preservation:} This term enforces a safety margin $T_I$ (set to 0) for oxygen concentration in void regions relative to accumulated polymer, ensuring that inhibition remains the dominant mechanism.
    \end{itemize}
    
    The threshold parameters $T_P$, $T_\text{OP}$, and $T_I$ are determined heuristically based on the energy partitioning between inhibitor depletion and polymerization propagation. These values encapsulate the ratio of energy consumed by inhibition versus that available for initiation. Specifically, inhibitor depletion consumes a fraction of $\frac{1.00}{1.62}$ of the total energy dose, leaving $\frac{0.62}{1.62}$ to drive polymerization.
    
    To accommodate spatially heterogeneous oxygen profiles, we simulate vertical oxygen concentration gradients for varying diffusion durations, as detailed in \autoref{sec:forwardmodel}. These concentrations are normalized against the baseline concentration at $\SI{40}{\celsius}$, $c_{\SI{40}{\celsius}} = \SI{6.421}{\mg\per\L}$. Consequently, a print initiated immediately post-preparation (\SI{0}{\minute}) assumes $c[\ce{O2}] \approx 1$, whereas prolonged storage (e.g., \SI{48}{\hour}) results in an initialized state of $c[\ce{O2}] \approx 2.28$, corresponding to the solubility ratio $c_{\SI{0}{\celsius}} / c_{\SI{40}{\celsius}}$.
    
\subsection{Numerical algorithm}
    Algorithm~\ref{alg:chem} describes the numerical procedure for the coupled differentiable ray-optical and photochemical optimization. This scheme adapts the methodology of \citet{Wechsler_Rizzo_Moser_2026} to explicitly model oxygen as the sole inhibiting species.
    
    Oxygen diffusion dynamics throughout the printing process are simulated using the Green's function for the diffusion equation:
    \begin{equation}
        K^{\Delta T}(x,y,z) = \frac{1}{(4\pi D \Delta T)^{3/2}} \exp\!\left(-\frac{x^2+y^2+z^2}{4D \Delta T}\right),
        \label{eq:kernel}
    \end{equation}
    where $D$ is the oxygen diffusion coefficient and $\Delta T$ is the simulation time step. We evaluate the diffusion operator numerically via a Fast Fourier Transform (FFT) convolution. To eliminate circular convolution artifacts characteristic of spectral methods, the simulation domain is padded spatially; given the compact support of the kernel, additional internal zero-padding or cropping is unnecessary.
    
    Light intensity distributions governing radical generation are quantified using the Dr.TVAM framework~\cite{Nicolet_Wechsler_Madrid-Wolff_Moser_Jakob_2024}, which integrates ray-optical phenomena such as refraction, reflection, scattering, and absorption. For a set of modulation patterns $P_j$ projected from angles $\alpha_j$, the total absorbed intensity $I(\mathbf r)$ is derived via the ray-optical operator $\mathcal{R}$:
    \begin{equation}
        I(\mathbf r) = \sum_{j=1}^{S} \mathcal{R}_{\alpha_{j}} P_j
        \label{eq:rayoptics}
    \end{equation}
    where $\alpha_j = \frac{j-1}{S} \cdot \SI{360}{\degree}$. The operator $\mathcal{R}$ maps projected intensities into the volumetric resin domain, accounting for refractive indices, attenuation coefficients, pixel dimensions, and vial geometry.
    
     Gradients required for the inverse design are computed via automatic differentiation. Within the algorithmic workflow, $c[\ce{O2}]$, $c[\ce{R}]$, and $c[\ce{P}^*]$ represent the concentrations of oxygen, radicals, and the effective polymerization dose, respectively.
    
    \begin{algorithm}[h]
    \caption{Surrogate algorithm with inhibition and diffusion}
    \label{alg:chem}
    \begin{algorithmic}[1]
    \State \textbf{Input:} Target design, time step $\Delta T$, total printing time $T_\text{exp}$, diffusion kernel $K^{\Delta T}$
    \State \textbf{Output:} Optimized TVAM patterns $\{P_j\}_{j=1}^K$
    \State $N_T \gets  T_\text{exp} / \Delta T$
    \Statex
    
    \For{optimization iteration $m = 1,2,\dots,M$}
        \State $c[\ce{P}^*] \gets 0$
        \State \textbf{Initialize oxygen field:}
        \State $c[\ce{O2}] \gets c[\ce{O2}]_0$
        \State \textbf{Compute absorbed intensity via Dr.TVAM:}
        \State Calculate volumetric light dose $I(\mathbf r)$ (Eq.~\ref{eq:rayoptics})
        \Statex
        \For{$n = 1,2,\dots,N_T$}
    
            \State \textbf{Radical generation:} 
            \State $c[\ce{R}] \gets I(\mathbf r) / N_T$ 
            \State \textbf{Oxygen quenching kinetics:} 
            \State $q_{\ce{O2}} \gets \min\!\left(c[\ce{R}],\, c[\ce{O2}]\right)$
            \State $c[\ce{R}] \gets c[\ce{R}] - q_{\ce{O2}}$
            \State $c[\ce{O2}] \gets c[\ce{O2}] - q_{\ce{O2}}$
            \State \textbf{Polymerization propagation:} 
            \State $c[\ce{P}^*] \gets c[\ce{P}^*] + c[\ce{R}]$
            \State \textbf{Diffusion step:}
            \State{\# $\ast$ \textit{denotes convolution}}
            \State $c[\ce{O2}] \gets c[\ce{O2}] \ast K^{\Delta T}$ (Eq.~\ref{eq:kernel})
        \EndFor
    
        \State Evaluate loss $\mathcal{L}\!\left(c[\ce{P}^*], c[\ce{O2}]\right)$ (Eq.~\ref{eq:loss})
        \State Update $\{P_j\}$ via L-BFGS optimization
    \EndFor
    \end{algorithmic}
    \end{algorithm}

\newpage
\section{Optimization results}
    \label{sec:opt_results}
\begin{figure}[H]
        \centering
        \includegraphics[width=1\linewidth]{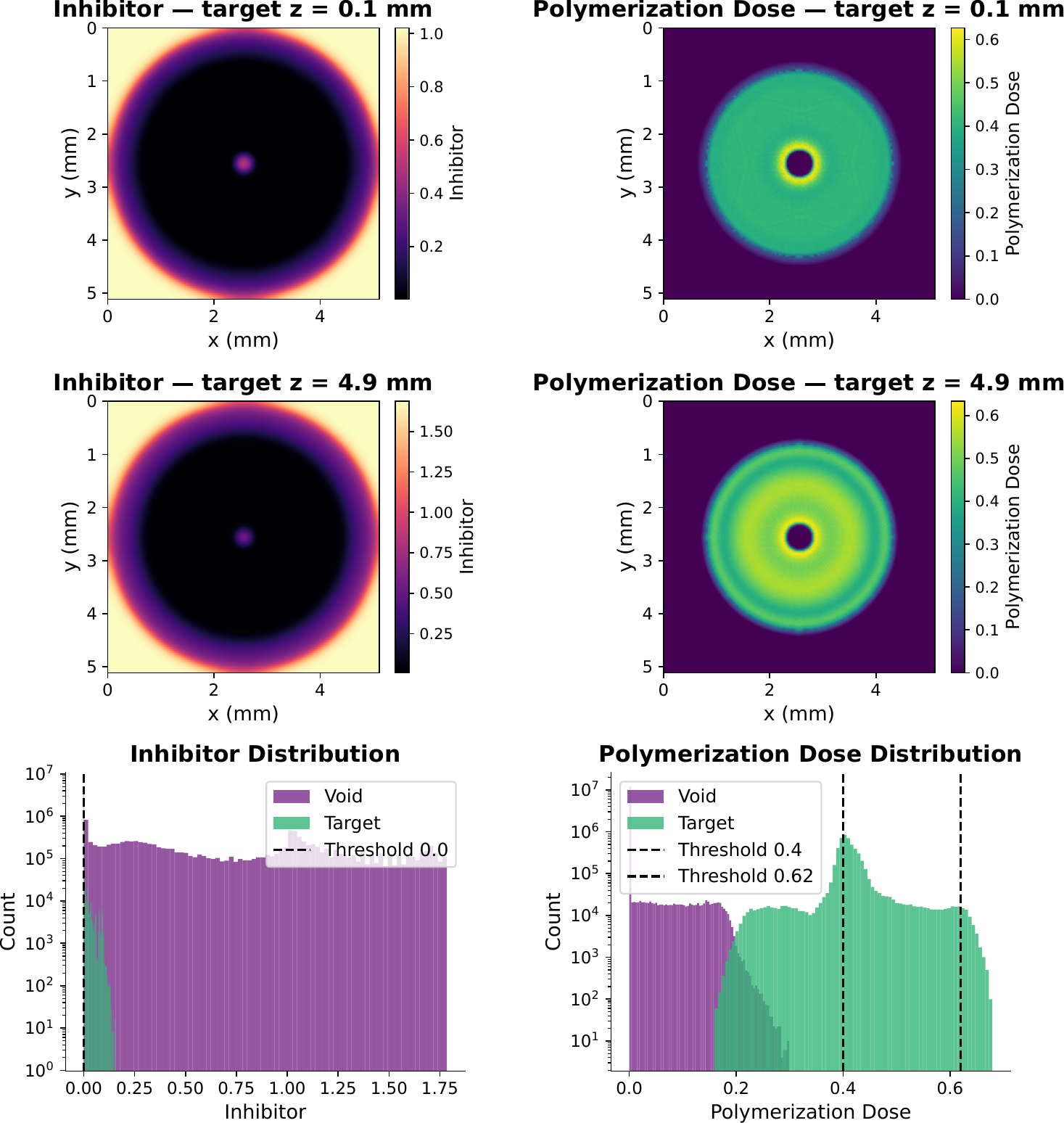}
        \caption{Optimization results for \SI{1}{\hour} resin preparation. Rows 1 and 2: Remaining inhibitor concentration (left) and polymerization dose (right) at $z=\SI{0.1}{\milli\meter}$ and $z=\SI{4.9}{\milli\meter}$. Row 3: Histograms of inhibitor and polymerization distributions plotted on a logarithmic scale.}
        \label{fig:optimization results}
    \end{figure}

    \autoref{fig:optimization results} presents the optimization results (40 iterations with L-BFGS) for a resin storage time of \SI{1}{\hour}. The figure visualizes the spatial distribution of the chemical species and the performance of the optimization.
    For all specified optimization parameters, see the source code repository.
    
    The first two rows display the remaining inhibitor concentration (left column) and the final polymerization dose (right column). The first row corresponds to a slice at $z = \SI{0.1}{\milli\meter}$ ($z=\SI{0}{\mm}$ corresponds to the bottom edge and $z=\SI{5}{\mm}$ to the top edge of the target), while the second row shows the slice at $z = \SI{4.9}{\milli\meter}$. A comparison of the two depths reveals that the inhibitor concentration is notably higher at $z = \SI{4.9}{\milli\meter}$. This occurs because oxygen had sufficient time to diffuse at this height before the print.
    
    The third row provides histograms for the inhibitor and polymerization levels within the target and void regions, plotted on a logarithmic scale. The results demonstrate that the optimization successfully separates the target and void distributions well. Perfect separation is rarely achievable for large threshold distances. However, numerically this result represents a high-fidelity print; the logarithmic scale visually overemphasizes low-probability regions that do not significantly affect the final structural quality.

    \newpage

\begin{figure}[H]
    \centering
    \includegraphics[width=1\linewidth]{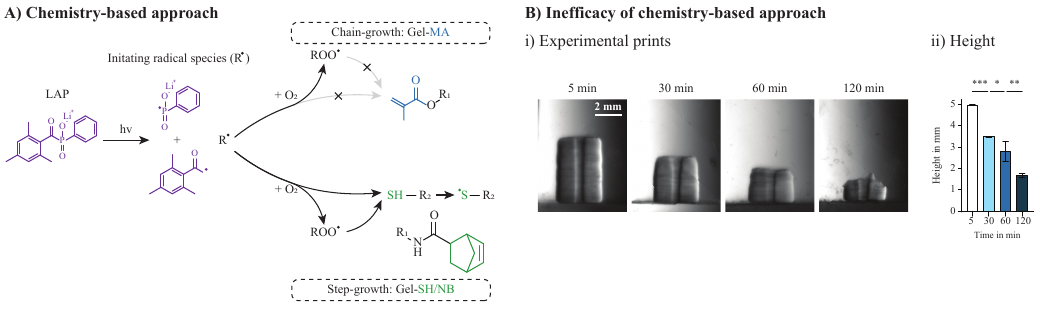}
    \caption{A) Schematic of chain-growth (Gel-MA) and step-growth (Gel-SH/NB) free-radical crosslinking mechanisms. While in (meth)acryloyl systems oxygen reacts with carbon-centered radicals to form unreactive peroxy radicals, in thiol-ene resins these peroxy radicals can still regenerate thiyl radicals and continue the crosslinking mechanism. B) Experimental prints (i) and heights (n = 3) (ii) showing that, although "insensitive to oxygen" also thiol-ene resins remain sensitive to spatial differences in dissolved oxygen concentration which still translate into textit{Pandoro effect}-related defects.}
    \label{fig:oxygen_insensitive}
\end{figure}

\begin{figure}[H]
    \centering
    \includegraphics[width=0.5\linewidth]{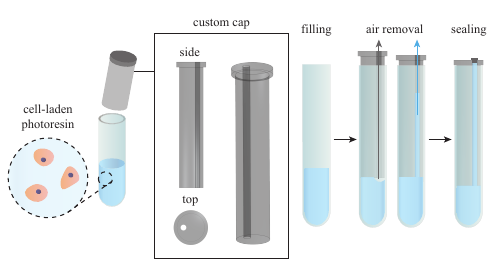}
    \caption{Illustration of custom cap designed to limit usage of cell-laden photoresin in engineering-based approach. By inserting the custom cap featuring a venting hole, air is first evacuated and then replaced by resin. The custom cap is then sealed to remove any air-resin interface.}
    \label{fig:customcap}
\end{figure}

\end{document}